\documentclass[conference]{IEEEtran}
\IEEEoverridecommandlockouts
\usepackage{cite}
\usepackage{amsmath,amssymb,amsfonts}
\usepackage{algorithmic}
\usepackage{graphicx}
\usepackage{textcomp}
\usepackage{xcolor}
\usepackage{hyperref}
\usepackage{comment}
\def\BibTeX{{\rm B\kern-.05em{\sc i\kern-.025em b}\kern-.08em
    T\kern-.1667em\lower.7ex\hbox{E}\kern-.125emX}}
\begin{document}

\title{Performance of a Hot Rb Vapour-Based Portable Quantum Memory\\}

\author{\IEEEauthorblockN{Kenneth Gregory}
\IEEEauthorblockA{\textit{Dept. of Electronics} \\
\textit{Carleton University}\\
Ottawa, Canada \\
KennyGregory@cmail.carleton.ca}
\and
\IEEEauthorblockN{2\textsuperscript{nd} Konrad Socha}
\IEEEauthorblockA{\textit{Dept. of Electronics} \\
\textit{Carleton University}\\
Ottawa, Canada \\
KonradSocha@cmail.carleton.ca}
\and
\IEEEauthorblockN{3\textsuperscript{rd} Khaled Mnaymneh}
\IEEEauthorblockA{\textit{Quantum and Nanotechnologies Research Centre} \\
\textit{National Research Council Canada}\\
Ottawa, Canada \\
khaled.mnaymneh@nrc-cnrc.gc.ca}
\and
\IEEEauthorblockN{4\textsuperscript{th} Connor Kupchak}
\IEEEauthorblockA{\textit{Dept. of Electronics} \\
\textit{Carleton University}\\
Ottawa, Canada \\
ConnorKupchak@cunet.carleton.ca}
}

\maketitle

\begin{abstract}
Quantum repeaters will be paramount to the success of the future quantum internet. Presently however, the losses incurred in fibre communications are unsustainable when fragile quantum states are transmitted over long distances. The most effective repeater nodes will require incorporation of a minimum of two quantum memory units (QMems) for successful operation. Therefore a network composed of numerous nodes will ideally require scalable repeaters that contain QMems that are portable, robust and do not require a high amount of overhead.  Here we investigate the performance of a commercial QMem based on a warm vapour of Rb atoms. Specifically, we characterize the coherence time, absolute efficiency, signal to noise ratio and bit error ratio  of this device over a range of vapour temperatures that are readily controllable.  For our system, we place these performance metrics into the same conventions as those established for current optical communication protocols. Development of QMem systems meeting these benchmarks will be a vital step for seamless integration of QMems systems into present-day communication architectures.

\end{abstract}


\section{Introduction}
The present-day internet is arguably one of the most powerful communication tools of the modern era, enabling high-speed global connectivity among other attributes. This capability is largely attributed to the extensive network of optical fibre lines that provide broadband communication to users worldwide. Successfully establishing these links requires the communications infrastructure to meet key metrics, including signal-to-noise ratio (SNR) and bit error ratio (BER). 

Looking ahead, these communication architectures are striving to attain fully guaranteed secure transmission with the establishment of the future quantum internet.  However, for the quantum internet to be viable, the chosen infrastructure must be able to transmit fragile quantum states over long distances. Optical transmission is inherently challenging due to the exponential loss of light as it propagates through optical fibres. Currently, fibre transmission is limited to a few tens to a few hundred kilometers.

In classical optical communication schemes, this loss is mitigated using optical amplifiers placed at intermediate locations. However, in quantum communications, these amplification schemes cannot be implemented due to the no-cloning theorem \cite{b1}. Therefore, quantum communication requires numerous quantum repeater nodes \cite{b2} to overcome propagation loss and transmit quantum optical information over usable distances.

For distances greater than 100 km, achieving suitable rates for information distribution requires the use of so-called quantum repeaters nodes. Additionally, these quantum repeaters must be equipped with robust quantum memory devices that are capable of reliably receiving and releasing quantum optical information on demand \cite{b3}. Any quantum network composed of memory-based repeater nodes will require twice the number of memory devices as there are established links. Therefore, implementing such a network will demand QMem devices that are scalable, economical, have low overhead requirements, and ideally operate near ambient temperatures.

Furthermore, to seamlessly integrate these devices into existing fibre-optic communication infrastructures, their performance must be evaluated using metrics common in classical optical communications. These include SNR and BER where the latter measures the ratio of incorrectly received data bits. In this study, we present an analysis of a first-of-its-kind, portable, and scalable QMem device designed for deployment as part of a widely distributed quantum repeater node. Specifically, we focus on a warm vapour of $^{87}$Rb atoms operating under conditions of Electromagnetically Induced Transparency (EIT) for the storage and retrieval of optical pulses \cite{b4}. 

\section{System Model and Problem Formulation}
\subsection{System Model}

Our investigation focuses on the performance of a newly available, commercial, rack-mountable QMem system (Qunnect Qu-Mem), which is based on a warm rubidium vapour cell. We probe this system using the D1 transition of $^{87}$Rb (795 nm) with a tunable diode laser (MogLabs CEF795).

Upon exiting the diode laser system, the 795 nm field is split into three channels: a low-power path for spectroscopy and laser locking, a path for probe field generation, and a path for the control (pump) field. The diode laser is locked onto the $|^5S_{1/2} F=2\rangle \leftrightarrow |^5P_{1/2} F'=1\rangle$ hyperfine transition, which serves as the control field. Laser locking is achieved using a locking system (Qunnect Qu-Lock) based on saturated absorption spectroscopy.

The probe field is generated using an electro-optical modulator (EOM) set to the $^{87}$Rb hyperfine ground state splitting of 6834.68 MHz detuned from the control field. The relevant sideband of the laser field exiting the EOM is spectrally and channel separated by a temperature-controlled etalon (LightMachinery Inc.) with a finesse of 35. Temporal modulation of the control and probe fields is achieved using separate 110 MHz Acoustic-Optical Modulators (AOMs) for each field. The probe field has a Gaussian temporal profile with a Full Width at Half Maximum (FWHM) of $1.05 ~\mu s$. We estimate that our classical probe field pulses contain about $\tilde 10^{8}$ photons.  For storage and retrieval operations, the control field is amplitude modulated on and off using a step function with an 80\% duty cycle. Both the control and probe fields are input into the QMem system via fibre ports.

Any quantum memory device used in a quantum repeater network must be indiscriminate to the degree of freedom used for qubit encoding. The QMem device under investigation can accept qubits encoded in any arbitrary polarization. It achieves this by separating the canonical polarization components into two distinct spatial paths using a polarization beam splitter (PBS) arranged in a Sagnac configuration.

The vertical rail of the probe field is rotated to horizontal polarization, so both rails of the probe field qubit are horizontal when passing through the vapour cell. A similar Sagnac and polarization rotation configuration is applied to the control field, ensuring that both components of the probe field are spatially overlapped with the two rails of an orthogonally (vertically) polarized control field. Before entering the vapour cell, the dual rails of the overlapped control-probe HV pairs are converted into circular polarizations to create a two-field configuration that benefits more from optical pumping \cite{b4}.

\begin{figure}
    \centering
    \includegraphics[scale=0.13]{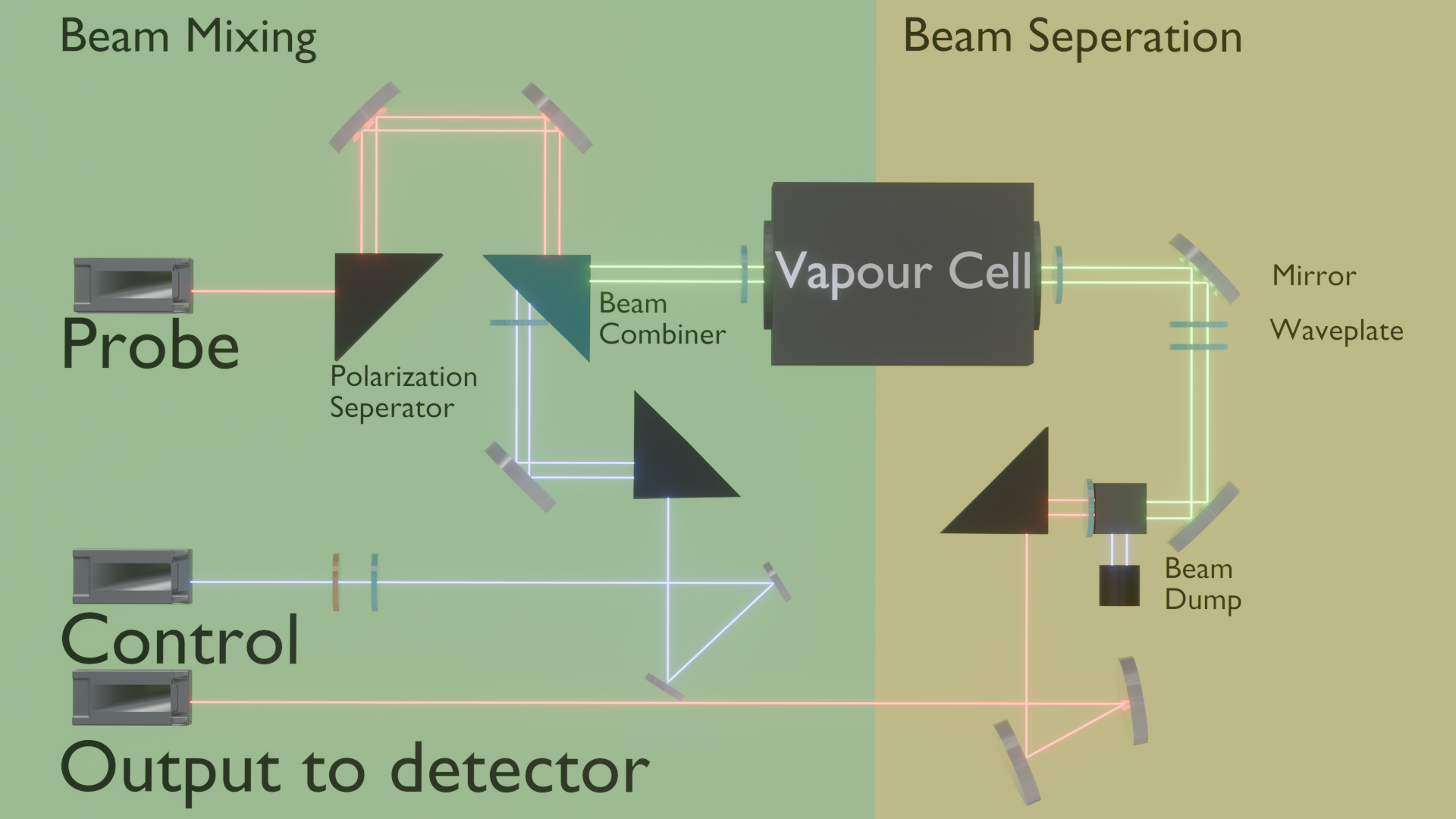}
    \caption{Schematic diagram of the interior of the QMem system. A single tunable diode laser generates both the control and probe beams. The probe beam is detuned from the control field by the $^{87}$Rb ground state splitting frequency using an electro-optical modulator (EOM). Both fields are coupled into separate input fibre ports of the QMem system. Before passing through the vapour cell, the probe and controlfields are spatially overlapped using a prism system.(Colour changed to green for clarity)}
    \label{fig:Schemdiagram}
\end{figure}

Upon exiting the vapour cell, the control  field is polarization-filtered using a Glan-Taylor calcite crystal. A second PBS-Sagnac system then recombines the probe rails to a single channel in the original input polarization state. After passing through the vapour cell, the output qubit is out coupled and sent to a photodetector for data collection via an oscilloscope. A schematic of the optical setup in the interior of the QMem system is shown in Fig.\ref{fig:Schemdiagram}.

\subsection{Problem Formulation}
The QMem system under investigation uses Electromagnetically Induced Transparency (EIT) for optical storage in a warm atomic vapor. In EIT, two non-degenerate atomic ground states are resonant with a common excited state \cite{b5}.  Transparency is achieved when the two optical dipoles driven by the laser fields interfere at the excited state, making the atoms transparent to both fields. This interference effect also alters the group index, trapping the atomic populations in the ground states \cite{b6}. By adiabatically modulating the intensity of the control field while the probe pulse is propagating through the cell, the pulse group velocity is reduced to zero, effectively storing the probe light pulse. The probe pulse can then be released on demand by reactivating the control field \cite{b7}.

The efficiency of the storage operation is quantified by the ratio of the time-integrated retrieved pulse to the time-integrated input pulse, as shown in Eq. \ref{eq:efficiency}:

\begin{equation}
    \eta = \frac{\int \left (S_{Pulse}-S_{Control} \right)\,dt}{\int \left(S_{Ref}-\zeta_{Off}\right)\,dt}\cdot 100\%.
    \label{eq:efficiency}
\end{equation}

\begin{figure}
    \centering
    \includegraphics[scale = 0.232]{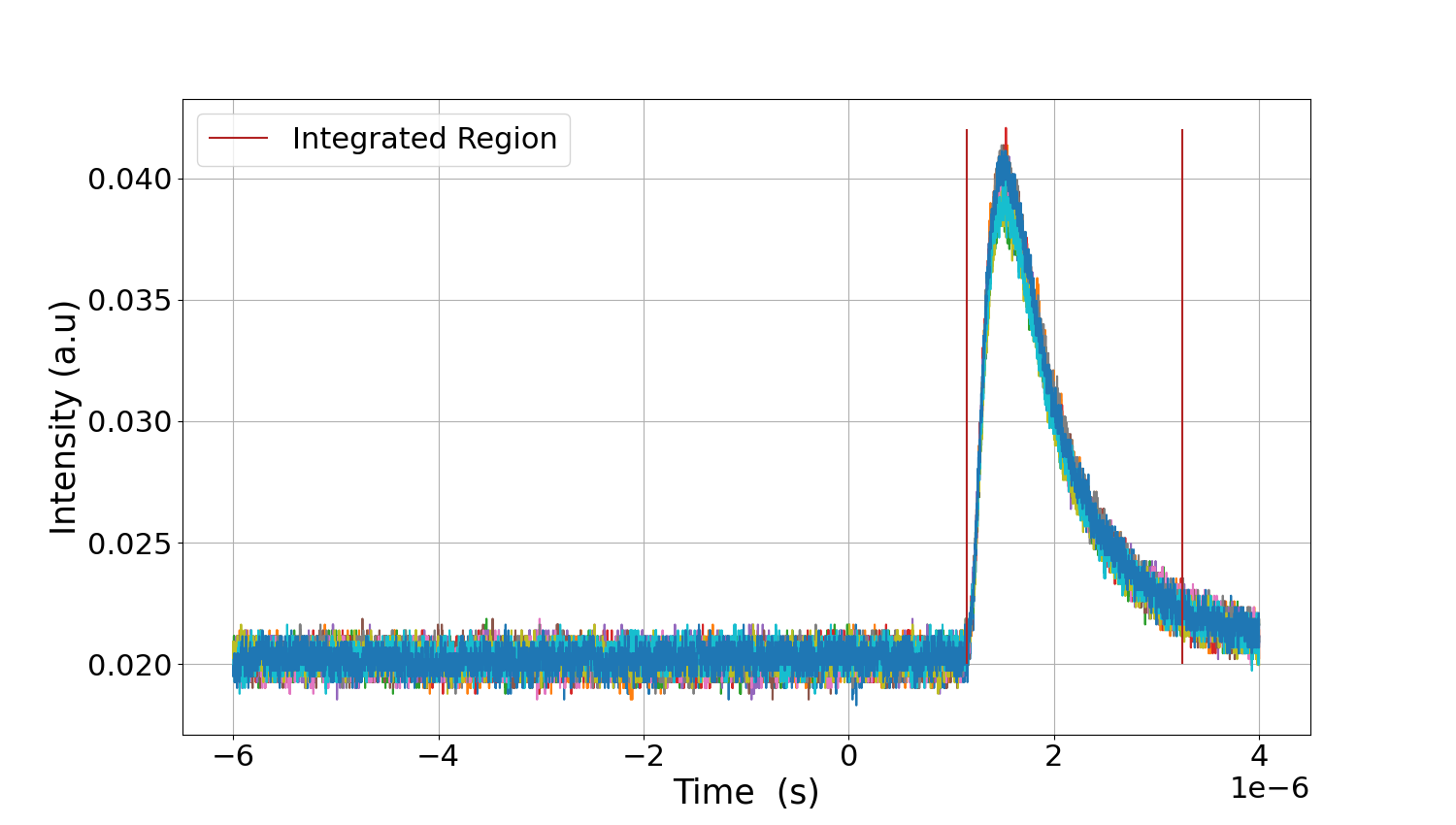}
    \caption{Sampling of retrieved probe pulse at a storage time of $50 ~\mu s$ and temperature of $70^{\circ}$C. The traces shown are not subject to any averaging.}
    \label{fig:Probe_Stability}
\end{figure}

The time integration of the probe field pulse yields the pulse energy. In our efficiency calculations, the temporal dynamics are obtained by averaging the traces from 65536 events for the input pulse, $S_{Ref}$  and 64 for the retrieved pulse, $S_{Pulse}$. This averaging helps to eliminate the effects of small amplitude fluctuations from the diode laser, as shown in Fig. \ref{fig:Probe_Stability}.  The transmission experienced by the probe pulses when propagating through the vapour cell system was found to be 0.75.

To account for noise and background contributions, we measure the background signals under storage and retrieval conditions in the absence of the probe field. This background trace is subtracted from the retrieved pulses to effectively remove the portion of the signal attributed to the control field, $S_{Control}$, and the detector offset, $\zeta_{Off}$. A similar operation is performed for the input pulse when in the absence of the control field.

\begin{figure}
    \centering
    \includegraphics[scale=0.232]{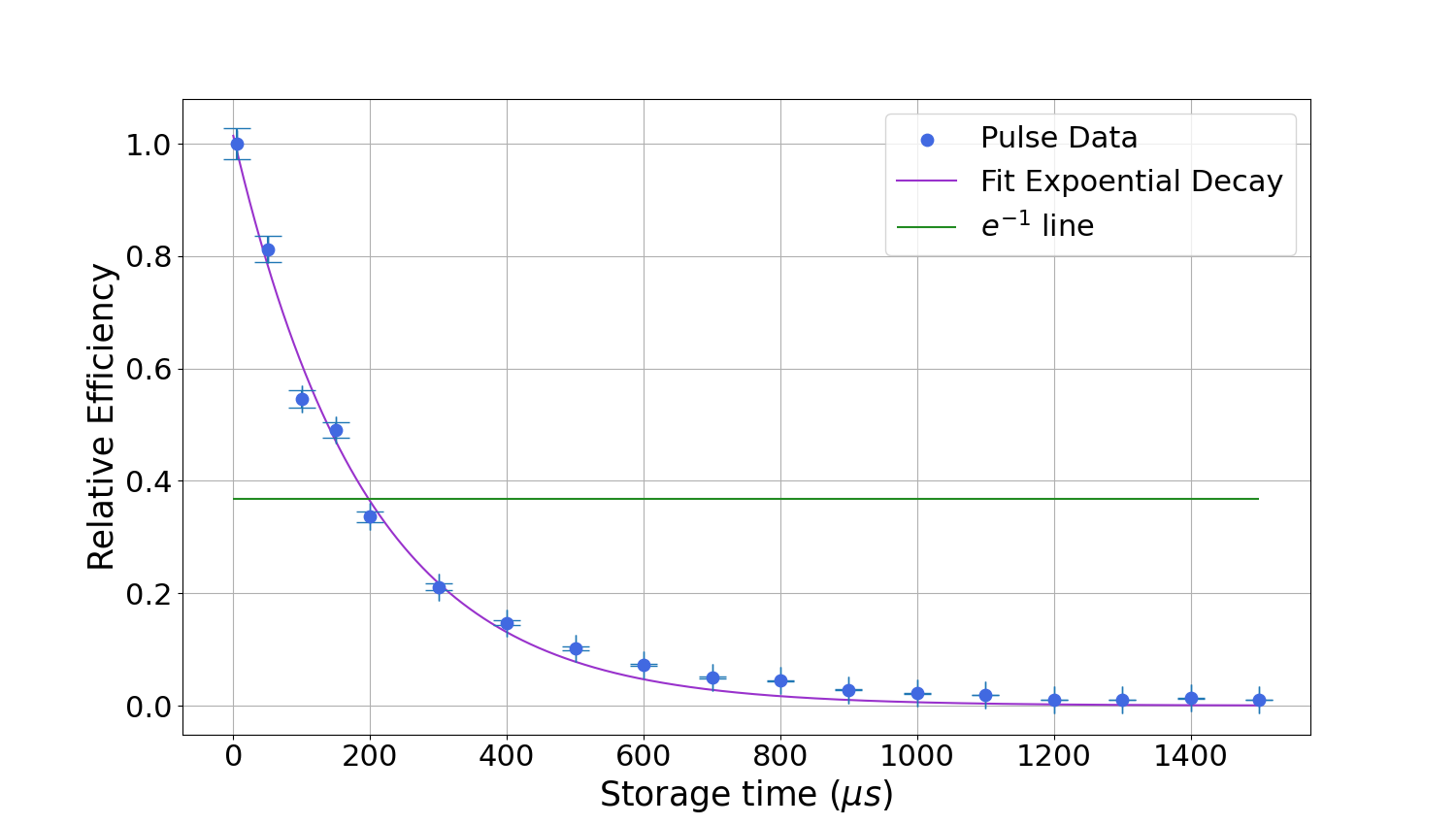}
    \caption{Storage efficiency as a function of storage time at a vapour temperature of $70^\circ$C. The coherence time is extracted from the fit to this curve.}
    \label{fig:70}
\end{figure}

By fitting these efficiency values to an exponential decay, we determine the coherence time of the optical storage by extracting the $1/e$ time, as shown in Fig. \ref{fig:70}. For long storage times, a significant amount additional classical offset is observed at retrieval. This is likely due to the decoherence of the ground state populations, which leads to the generation of photons that are incoherent with those of the stored probe fields. This classical offset is substracted from the output signals to achieve a more accurate fit of the coherence time.

In classical communications, distinguishing the information encoded in a received signal from background noise is crucial. The ability to differentiate between signal and noise is quantified by the SNR. A high SNR is important in both classical and quantum communications as it enables higher bandwidth operations and supports more complex modulation schemes. High SNR results in fewer errors, better data transmission quality, and improved signal integrity, enhancing overall system performance.  Both classical and quantum communication systems rely on signal processing and its relationship to SNR. However, in quantum systems, maintaining the quantum properties of the transmitted signal is an additional requirement. A low SNR in classical systems primarily affects data transmission quality, in quantum systems, it will additionally impact the security and quantum nature of the signals being transmitted.

In this investigation, we focus on SNR values within the classical regime. Future studies will explore how these classical SNR values translate to the quantum regime, as SNR is closely related to the fidelity of the quantum state \cite{b13}. For a memory system, fidelity measures how well the input and output quantum states match. Investigating our QMem system in the quantum regime will require advanced filtering systems to eliminate noise from scattering processes and probe field leakage \cite{b14}.  Here, to quantify SNR we average over a number of events (traces) denoted by $N$. The detector offset $\zeta_{Off}$ is subtracted from each trace, and the time-averaged signal is obtained by integrating over the region of interest~$T_{ROI}$.

 To determine the noise, we similarly subtract the detector offset and calculate the root mean square (RMS) of the detector fluctuations, $\sigma_{RMS,n}$. This value represents the noise inherent in the probe signals. The SNR is then calculated using Eq..
\ref{eq:SNR} as shown:
\begin{equation}
    SNR = \frac{1}{N}\sum_{n=1}^{N}\left(\frac{\frac{1}{T_{ROI}}\int_{T_{ROI}} \left(S_n-\zeta_{Off}\right)\,dt }{\sigma_{RMS,n}}\right).
    \label{eq:SNR}
\end{equation}
Quantifying the SNR value for the QMem systems is necessary to determine the BER capabilities of our QMem channel. These metrics allow us to compare the performance of the QMem within the context of classical communication protocols.

\begin{figure}
    \centering
    \includegraphics[scale=0.232]{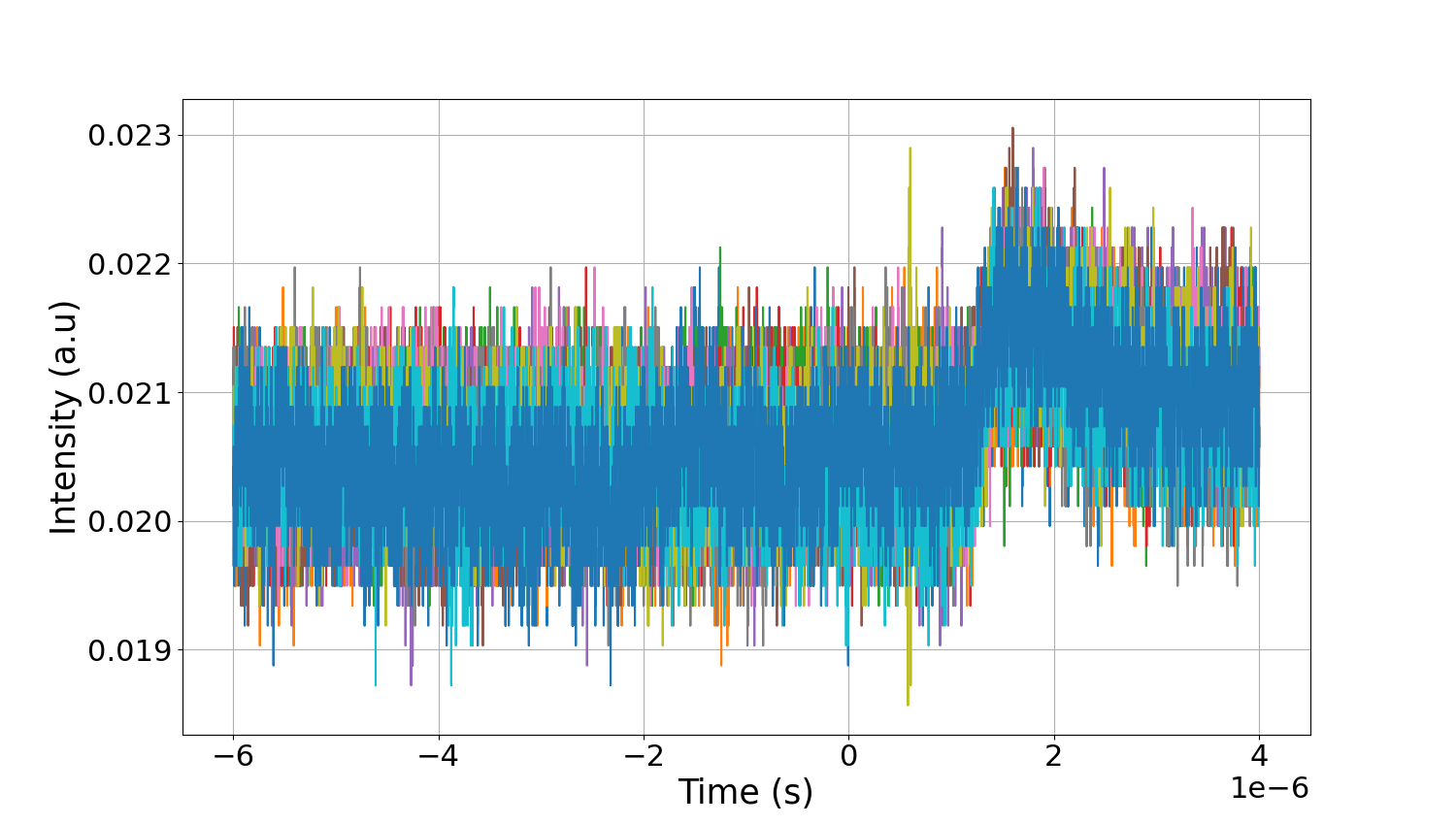}
    \caption{Sample of noise traces for a when the control field is switched on after $1500 ~\mu s$. In the data shown, the probe field is not present and the effect of atomic decoherence between the atoms is apparent starting at $1.25~\mu s$}
    \label{fig:Noise}
\end{figure}

\section{Results and Discussion}

\subsection{Results}

\begin{figure}
    \centering
    \includegraphics[scale=0.232]{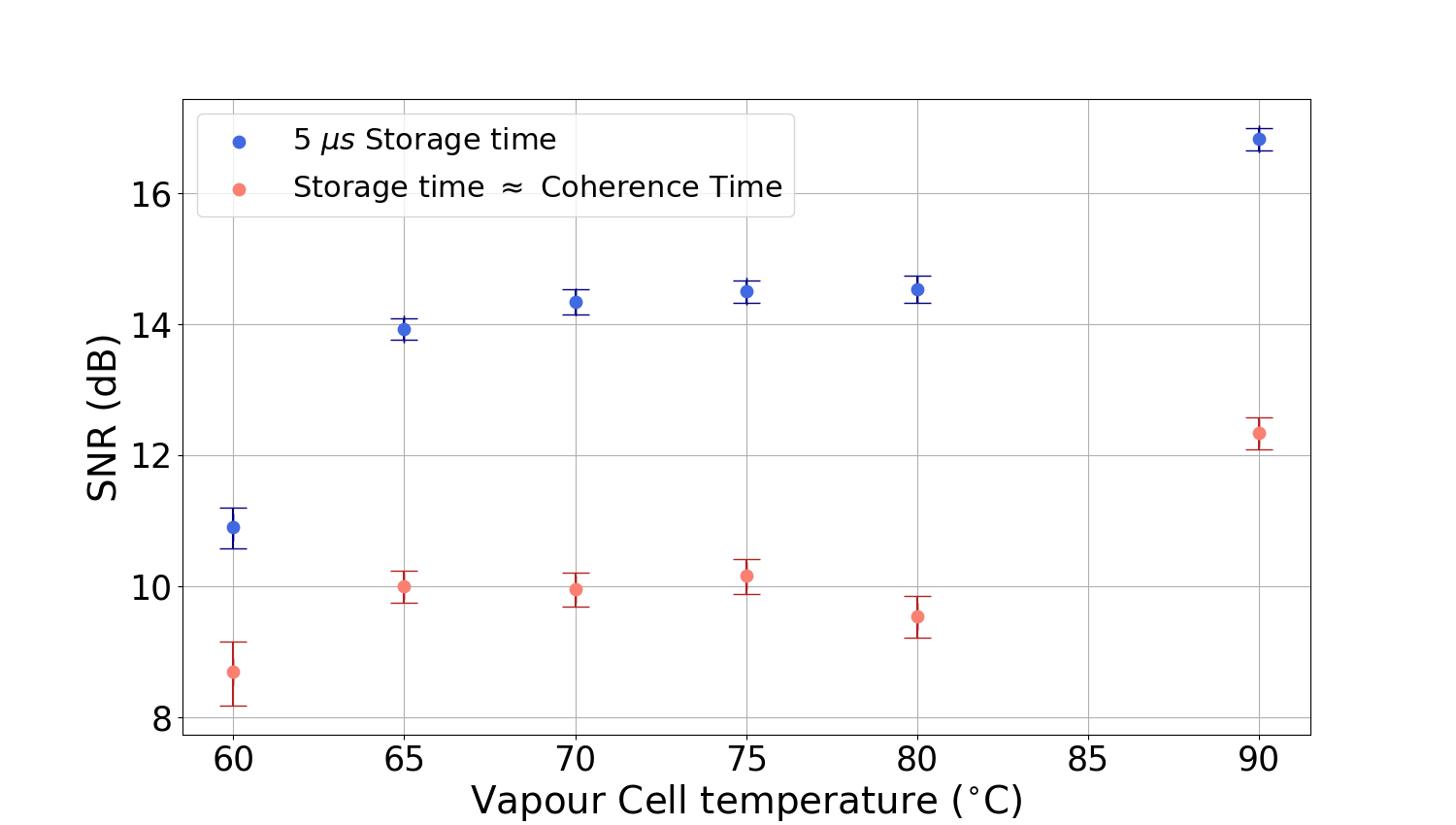}
    \caption{The signal to noise ratio over $60^\circ$C - $90^\circ$C temperature range. The SNR values for both $5~\mu s$ and for storage times that are near the $1/e$ time for a given temperature.}
    \label{fig:SNR_Temp}
\end{figure}

\begin{figure}
    \centering
    \includegraphics[scale=0.223]{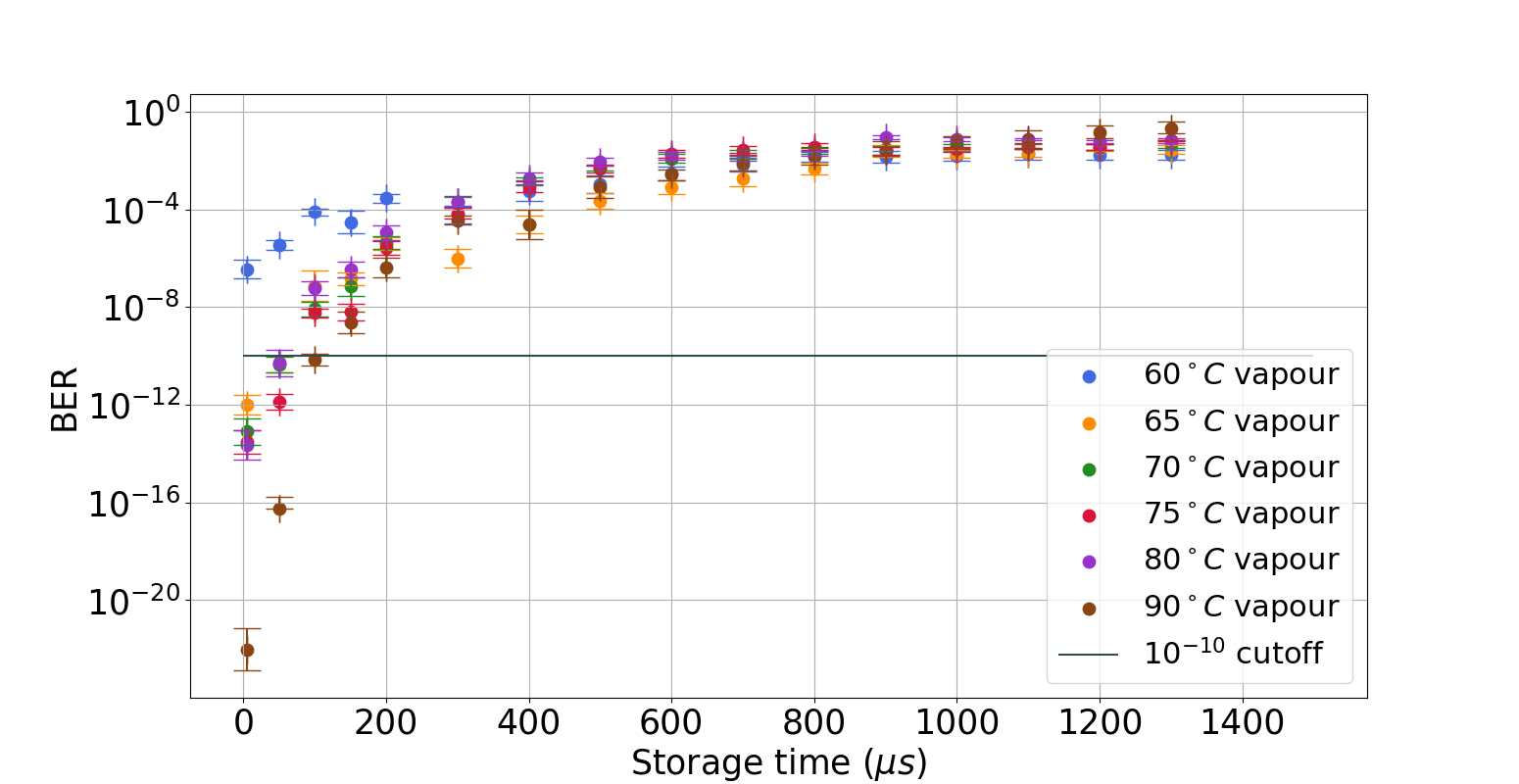}
    \caption{The plot shows the BER versus storage time at the temperatures investigated in the QMem system. The BER values were calculated using the measured SNR values.}
    \label{fig:BER}
\end{figure}

First we investigated the coherence lifetime and of the QMem system across a range of temperatures as shown in Fig. \ref{Fig:coh}. In general, the coherence time of the optical storage decreased as the vapour cell temperature increased. Next for the efficiency, we observed a non-monotonic behavior, with a peak value near $70^{\circ}$C as shown in Fig. \ref{Efficiency.png}. Next we studied the SNR values over the same temperature range as shown in Fig. \ref{fig:SNR_Temp} The maximum SNR value was found to occurr at a vapour temperature of $90^\circ$C, which was the highest temperature tested in this study.  Lastly, we analyzed the BER as a function of storage time for different temperatures as shown in Fig. \ref{fig:BER}. The BER values were calculated based on the measured SNR values, following the method described in \cite{b8}:

\begin{figure}
    \centering
    \includegraphics[scale=0.232]{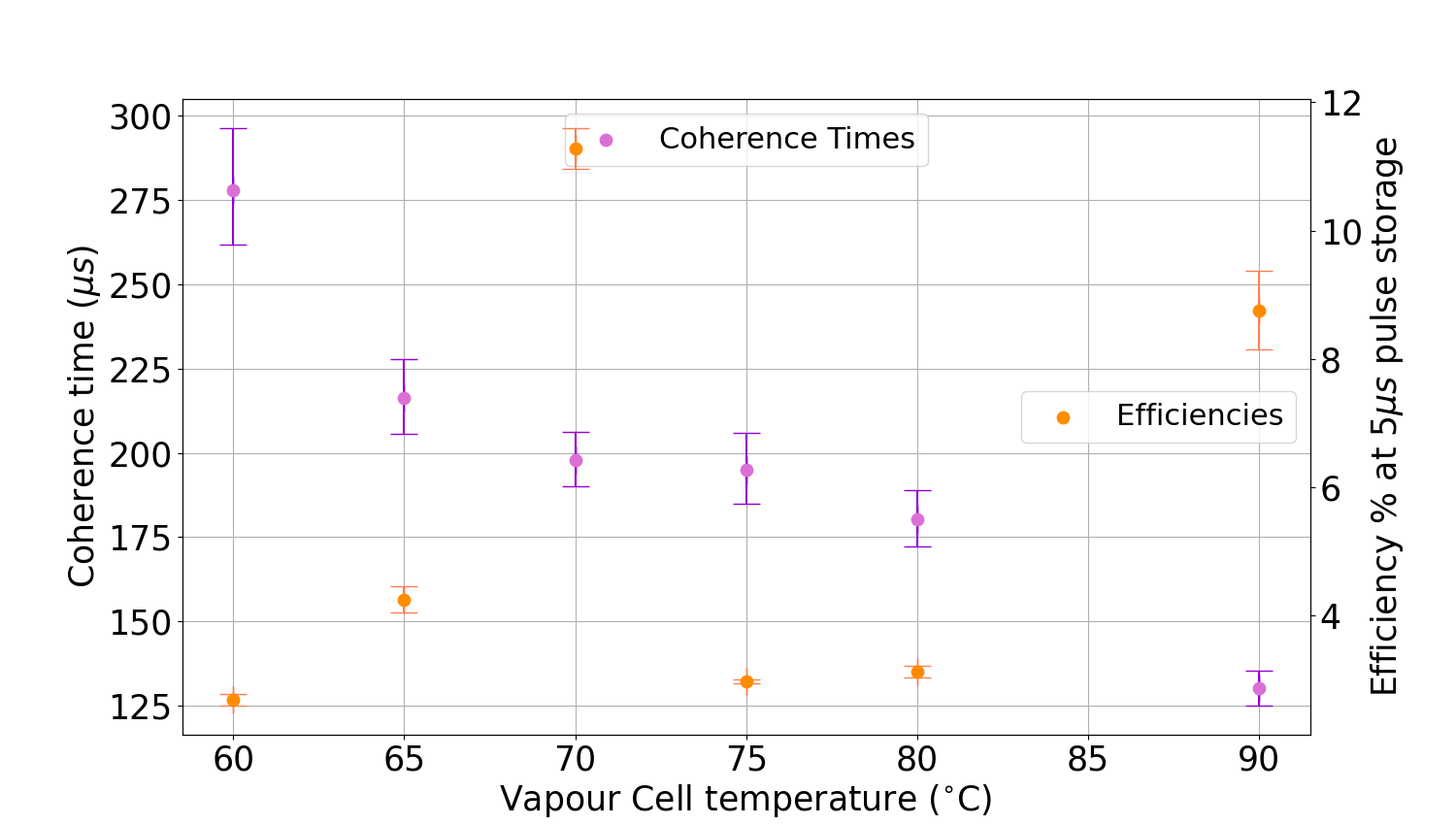}
    \caption{The Coherence time and efficiency of optical storage over a temperatures ranging from  $60^\circ$C - $90^\circ$C.  The error bars are determined by statistical analysis over multiple events.}\label{Fig:coh}
\end{figure}

\begin{figure}
    \centering
    \includegraphics[width=\linewidth]{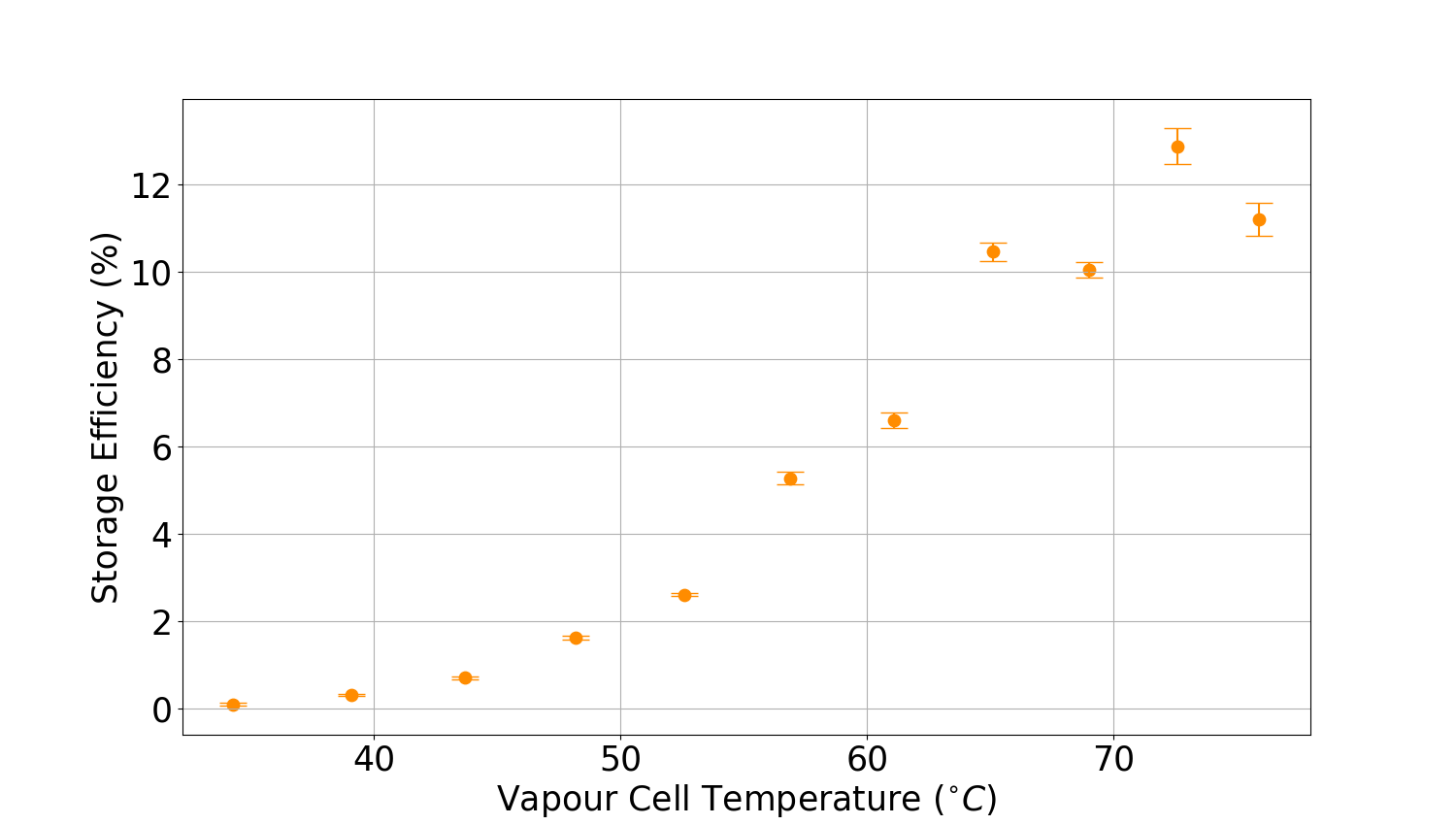}
    \caption{A second investigation into the temperature dependence of storage efficiency. A larger range of temperatures is tested, and the plot is calibrated to the sensor reading of the vapour cell rather than the set-point temp. The efficiency is defined by the ratio of the integrals of a retrieved $5~\mu s$ pulse and a reference pulse when no vapour cell is present. The error bars are determined by statistical analysis over multiple events.}
    \label{fig:Eff}
\end{figure}

\begin{equation}
BER=Q(\sqrt{SNR}),
\end{equation}
where the Q-function can be found by
\begin{equation}
Q(x)=\frac{1}{\sqrt{2\pi}}\int^\infty_x e^{-y^2/2}dy.
\end{equation}

\subsection{Discussion} 
In a perfect quantum repeater, the memory is able to store a photon for an indefinite time as the other nodes establish entangled links. A realized quantum memory will not be able to store a signal forever as decoherence processes slowly deteriorate the signal strength. The nominal measurement of this decay is the coherence time, when the signal power has reduce to $1/e$ of the original.  These decoherence processes could the effects of fast moving atoms moving in and out of the active beam area. In a vapour system, the molecular speed is determined by their kinetic energy, given by the equation $\frac{3}{2} kT = \frac{1}{2}mv^2$, where $T$ is the temperature in Kelvin, $m$ is the atomic mass, $k$ is the Boltzmann constant, and $v$ is the average atomic speed. For a system going from 353K to 383K, there is a resulting 4\% increase in the speed, so an increase decoherence due to atomic movement is expected to be minimal.

Our data indicates the presence of other decoherence mechanisms at higher temperatures, including ground state population exchange due to atomic collisions and increased sensitivity to beam vector fluctuations caused by temperature gradients in the setup \cite{b4}. In our datasets, temperature data was recorded from highest to lowest due to differences in heating and cooling times.  One would expect the storage efficiency to decay exponentially to zero. However, atomic noise caused by population exchange results in a non-zero signal at long storage times, affecting the true decay rate. When correcting for this noise offset, the coherence time is found to be similar to other studies \cite{b4}. For example, at $60^\circ$C, the coherence time of our system was found to be $278 \pm 18 ~\mu s$.

For long-distance quantum communications, the usable memory storage time should align with the time required to establish a connection between nodes and transmit a classical signal back to the photon source, confirming that a successful link has been made. 


Regarding storage efficiency, we observe an increase with higher temperatures, as shown in Fig. \ref{fig:Eff}. This increase is expected because higher temperatures correlate with an increase in optical depth and a higher probability of atom-photon interactions. However, efficiency begins to decrease at higher temperatures, likely due to the re-absorption of the probe field before it exits the vapour cell. This behavior is similar to what was observed in Ref. \cite{b4}.

The values of SNR and BER are closely related; SNR determines if a detector can distinguish between receiving a signal, or if the response is from detector noise. A bit error comes when the detector either fails to respond to an incoming photon, or triggers off of a noise event as if a photon had arrived. An increase in SNR corresponds to a decrease in the probability of a bit error.  In our study, we found the optimal operating conditions for SNR and BER at a cell temperature of $90^\circ$C and a storage time of $5~\mu s$. The best conditions were defined as exhibiting a BER below the threshold value of $10^{-10}$ for the longest range of storage times. With the exception of $60^\circ$C, all the temperatures examined had BER values below the threshold for storage times up to $50~\mu s$.

Classical communication protocols require an SNR between 10 dB and 25 dB \cite{b9}-\cite{b11}. At a storage time of $5 ~\mu s$, our system achieves SNR values ranging from 10.9 dB to 16.8 dB, which falls within the stated range. For storage times near the measured $1/e$ coherence time at a particular temperature, the SNR ranges from 8.7 dB to 12.3 dB.  Hence, our system yields the SNR values necessary for to serve as a classical optical channel for storage times between $100-200 ~\mu s$.

Lastly, the expected BER can be determined from our measured SNR values at different storage times and vapour cell temperatures. The standard for acceptable BER in telecommunication systems is generally considered to be $10^{-9}$ \cite{b12}. As shown in Fig. \ref{fig:BER}, the QMem system exhibits BER values below $10^{-10}$ for storage times less than $<100~\mu s$ and temperatures above $60^\circ$ C. Therefore, under these conditions, our QMem system meets the benchmarks of classical telecommunication networks. Extending these metrics to the quantum level is not directly possible for the SNR and BER calculations put forward here; they are equipment dependant measurements, and the equipment does not have the required sensitivity to make quantum measures. On the other hand, the coherence time and efficiency are both macroscopic quantities of the Rb vapour in the cell, which are true regardless of photon count.

It is important to emphasize that our QMem system operates in the near-infrared regime at a wavelength of 795~nm. For this system to be viable in optical communication infrastructures, the photons will need to undergo wavelength conversion. Fortunately, rubidium exhibits a two-photon resonance that allows for the conversion of near-infrared photons to telecommunication wavelengths using a four-wave mixing diamond scheme \cite{b15,b16}.

\section{Conclusion}

We have presented an investigation into a portable QMem system based on Rb vapour that operates at near room temperature. These portable devices will be essential for realizing scalable quantum repeater networks in future quantum internet architectures. Specifically, in our QMem system, we characterized the coherence time, storage efficiency, SNR and BER across a range of easily adjustable temperatures.

We found the optimal coherence time at $60^\circ$C. The highest efficiency and SNR values were observed at higher temperatures of $70^\circ$C and $90^\circ$C, respectively. This suggests an optimal operational window for these QMem systems based on the desired performance characteristic. The SNR and BER values of the QMem systems presented here were calculated and compared directly to the accepted SNR and BER levels used in classical communication protocols. Our system demonstrated the capability of achieving these standardized benchmarks at certain temperatures and short storage times. Placing the metrics of our QMem system in this context is a key step toward integrating this technology into existing and future communication frameworks. Future studies will focus on improving these SNR and BER values for the quantum regime and translating them into secret key rates. Finally, entangled photon rates will be found as a function of storage efficiency and pulse width to determine the feasibility of implementation into quantum networks.

\section*{Acknowledgment}

The authors acknowledge funding from the NSERC Discovery Program and CFI-JELF. All authors also wish to thank the Qunnect team for all their help with troubleshooting, fruitful discussions and input.

\end{document}